\documentclass[aps,showpacs]{revtex4}
\usepackage{amsmath}
\usepackage{amssymb}
\newcommand{\bm}[1]{\mbox{\boldmath$#1$}}

\begin{document}
\title{Nonlinear Fluid Dynamics Description of non-Newtonian Fluids\\[0.1cm]}
\author{Harald Pleiner$^1$, Mario Liu$^2$, and Helmut R. Brand$^3$}
\affiliation{\vspace{0.2cm} $^1$Max-Planck-Institut f\"ur Polymerforschung, 55021 Mainz, Germany
 \\ 
$^2$Institut f\"ur Theoretische Physik, Universit\"at T\"ubingen, 72676 T\"ubingen, Germany\\
$^3$Theoretische Physik III, Universit\"at Bayreuth, 95440 Bayreuth, Germany}

\received{6 Janaury 2004}
\begin{abstract} 
\ \\[-0.2cm] Nonlinear hydrodynamic equations 
for visco-elastic media are discussed. We
 start from the recently derived fully hydrodynamic nonlinear 
description of permanent elasticity that utilizes the (Eulerian) 
strain tensor. The reversible quadratic nonlinearities in 
the strain tensor dynamics are of the 'lower convected' type, 
unambiguously. Replacing the (often neglected) strain diffusion 
by a relaxation of the strain as a minimal ingredient, a generalized
hydrodynamic description of viscoelasticity is obtained. 
This can be used to get a
 nonlinear dynamic equation for the stress tensor (sometimes called 
constitutive equation) in terms of a power series in the variables. 
The form of this equation and in particular the form of the nonlinear 
convective term 
is not universal but depends on various material parameters. A comparison with 
existing phenomenological models is given. In 
particular we discuss how these ad-hoc models fit into the hydrodynamic
description and where the various non-Newtonian contributions are coming from.
\end{abstract}

{\it Rheologica Acta} {\bf xx}, xxx (2004); DOI 10.1007/s00397-004-0365-8
\vskip 0.6cm

 \maketitle

\section{Introduction \label{intro}}

Hydrodynamics is a well established field to describe macroscopically simple 
fluids by means of the Navier-Stokes-, continuity, 
and heat conduction equations. However, it applies also to 
more complex fluids that are fully characterized by conservation laws 
and broken symmetries. This more general hydrodynamic method has been 
established in the 60s \cite{bogoljubov,kadanoff,hohenberg} and 
applied e.g. to superfluids \cite{khalatnikov} and liquid 
crystals \cite{MPP}. It is based on (the Gibbsian formulation of) 
thermodynamics \cite{callen,reichl}, symmetries and well-founded 
physical principles \cite{forster}. A detailed description of this 
method can be found in \cite{MPP,PB1}. Somewhat related approaches 
have been used for liquid crystals \cite{lee,leslie,ericksen,hess} 
and more generally in \cite{grmela2,beris}.

On the other hand, non-Newtonian fluids are believed to show 
non-universal behavior and a host of different empirical models have 
been proposed \cite{oldroyd,coleman,truesdell,giesekus,bird,larson} 
to cope with the flow rheology of such substances. Typically these 
models are formulated as generalizations of the linear, Newtonian 
relation between stress and deformational flow allowing for additional 
time derivatives and nonlinearities. They are tailored to accommodate 
empiric findings or are based on principles \cite{bird} that are 
ad-hoc and generally insufficient. 

Quite recently we have derived a nonlinear hydrodynamic description 
of elastic media \cite{temmen,p1} that is based on first principles, only, making use of thermostatics, linear irreversible thermodynamics, symmetries and broken symmetries, and invariance principles. It
has been confirmed within 
the GENERIC formalism \cite{grmela}. Allowing in this hydrodynamic 
description the strains to relax (and not only to diffuse) a generalized 
hydrodynamic description of nonlinear viscoelasticity is obtained in 
terms of a dynamic equation for the strain tensor \cite{temmen,p1}. 
 After a thorough exposition of this formulation (Sec.\ref{straindescr}) 
we transform it into a description in terms of a dynamic equation 
for the stress tensor (Sec.\ref{stressdescr}). This can only be done 
approximately in the form of a power expansion in the variables. 
Up to second order a formulation is obtained that can directly be compared 
with many of the empirical models proposed to describe non-Newtonian
rheology. The comparison (Sec.\ref{compar}) reveals possible 
inconsistencies and connects the various ad-hoc additions of those 
models with physical relevant processes, like strain relaxation, 
elasticity and viscosity. In two appendices we sketch possible 
extensions of the hydrodynamic description of viscoelasticity.

A comparison with recent constitutive equations that refer to specific 
microscopic variables and processes, like convective constraint release 
\cite{wagner,marrucci,mcleish}, will not be done in the present paper. 
Here we rather concentrate on the simplest generalized 
hydrodynamic description of non-Newtonian rheology in terms of a relaxing strain field, while
a detailed comparison with those theories
requires the use of additional relaxing fields.

\section{Strain tensor description \label{straindescr}}

In this section we review the hydrodynamic description of nonlinear 
elasticity and its generalization to viscoelasticity in terms of 
the Eulerian strain tensor $U_{ij}$. 
We start with a short reminder of its definition.
In a stress-free elastic body, we consider a point
at the (initial) coordinate $\bf a$. As the body is
displaced, rotated, compressed and sheared, the
given point is displaced to $\bf r$.
Since all points of the body have a unique pair of
$\bf a$ and $\bf r$, the function $\bf r(a)$ is
unique and invertible, the result of which we denote
as $\bf a(r)$. 
Describing physics in local terms, in particular, the
state variables characterizing the elastic body are taken as functions of the
real space coordinates $\bf r$ rather than of the
reference space coordinates $\bf a$. Therefore we choose $\bm{a}(\bm{r})$ rather than $\bm{r}(\bm{a})$ as the starting point to construct the (Eulerian) strain tensor
$U_{ij}=\textstyle\frac{1}{2}[\delta_{ij}-(\nabla_{j}
a_k) (\nabla_{i} a_k)]$, which can be written in the more familiar form
$U_{ij}=\textstyle\frac{1}{2}[\nabla_{j}
u_i + \nabla_{i} u_j - (\nabla_{i} u_k)(\nabla_{j} u_k)]$ using the displacement field $u_i(\bm{r}) = r_i - a_i(\bm{r})$.
In the Lagrangian description, where all fields are function of the initial coordinates rather than the local ones,  the (Langrangian) strain tensor is $U_{ij}^L=\textstyle\frac{1}{2}[(\partial r_k / \partial a_j)(\partial r_k / \partial a_i)-\delta_{ij}]$ or $U^L_{ij}= \textstyle \frac{1}{2}
[\partial u_i/\partial a_j+ \partial u_j/\partial a_i+
(\partial u_k/\partial a_i)\cdot(\partial u_k/\partial
a_j)]$, with $u_i(\bm{a})=r_i(\bm{a})-a_i$.
Since the dynamics for $\bm{a}$ is simply $d\bm{a}/dt =0$ in the absence of phenomenological currents, the dynamics for the Eulerian strain tensor reads \cite{temmen,p1}
\begin{equation} \label{udyn}
 \frac{d}{dt}U_{ij} - A_{ij} + U_{ki} \nabla_j v_k + 
U_{kj} \nabla_i v_k =  X_{ij}^{(ph)}
\end{equation}
with $d/dt \equiv \partial/\partial t + \bm{v \cdot \nabla}$, 
where $v_i$ is the velocity and $A_{ij} = \tfrac{1}{2}( 
\nabla_i v_j + \nabla_j v_i)$ its symmetrized gradient 
describing deformational flow. This equation contains, 
apart from the transport derivative $\sim \bm{v \cdot \nabla}$ due 
to Galilean translational invariance, a linear 
and a nonlinear coupling to flow. The former ($\sim A_{ij}$) 
reflects the spontaneously broken translational symmetry 
in elastic media (in an  equation for the displacement vector the 
equivalent term subtracts the solid body translations, which must 
not change the elastic state of the fluid). The nonlinear one  
is of the lower convected type and results from the freedom to 
choose the orientation of a material-fixed frame independently from 
that of the laboratory frame \cite{temmen, p1}. For a Lagrangian 
strain tensor the upper convected derivative  would be obtained. 
On the right hand side there is the phenomenological (quasi-)current 
$X_{ij}^{(ph)}$, which describes, in the case of permanent elasticity, 
purely diffusional processes. 
In that case Eq.(\ref{udyn}), when combined with the dynamic equations for the
other hydrodynamic variables, the density, $\rho$,
the energy density, $\epsilon$, and 
the density of linear momentum, $g_i$, gives rise to three additional 
truly hydrodynamic modes, i.e. the two doubly degenerated transverse 
sound modes (compared to the two vorticity diffusions of simple 
fluids) and vacancy diffusion \cite{MPP}. 
For decaying strains in viscoelastic media  $X_{ij}^{(ph)}$ 
contains relaxational processes (see below). The strain tensor 
$U_{ij}$ is symmetric, in order to exclude solid body rotations, 
since the latter must not change the elastic state.

The second relevant dynamic equation is the momentum balance or generalized
Navier-Stokes equation
\begin{equation} \label{vdyn}
\rho  \frac{d}{dt} v_i  + \nabla_i p + \nabla_j \sigma_{ij} =0
\end{equation}
with the isotropic pressure $p$ and the stress tensor $\sigma_{ij}$, 
which is of the form
\begin{equation} \label{stress}
\sigma_{ij} = - \Psi_{ij} + \Psi_{ki} U_{jk} 
+ \Psi_{kj} U_{ik} + \sigma_{ij}^{(ph)}
\end{equation}
Here $\Psi_{ij}$ is the {\em elastic} stress tensor, the thermodynamic 
conjugate to the strain tensor. The linear and nonlinear 
$\Psi_{ij}$ contributions are the counterterms to the linear 
and nonlinear flow contributions in Eq. (\ref{udyn}) that are necessary 
to cancel any contribution to the entropy production \cite{PB1}, 
since all these terms are reversible. Thus, these parts of the 
stress tensor are completely fixed by general physical principles. 
The phenomenological part of the stress tensor is contained 
in $\sigma_{ij}^{(ph)}$ (see below) and describes in the simplest 
form Newtonian viscosity. The stress tensor is symmetric 
(or can be made so) in order to guarantee angular momentum 
conservation \cite{MPP}. 

Throughout this paper we will assume incompressibility that reduces 
the continuity or mass conservation equation to ${\rm div} \bm{v} =0$. 
In addition we neglect the thermal degree of freedom and, thus, 
the heat conduction equation.

Phenomenological or material specific properties enter our equations 
on three different occasions, two are dynamic and one is static 
in nature. For the former we stay within the reach of well-founded 
linear irreversible thermodynamics \cite{dGM}, i.e. we assume a 
linear relation between the "forces" $\Psi_{kl}$ and $A_{kl}$ 
and the "fluxes" $X_{ij}^{(ph)}$ and $\sigma_{ij}^{(ph)}$
\begin{eqnarray} \label{relax}
X_{ij}^{(ph)} &=&-\alpha_{ijkl} \Psi_{kl}
\\ \label{visc}
\sigma_{ij}^{(ph)} &=& - \nu_{ijkl} A_{kl}
\end{eqnarray}
describing strain relaxation and viscosity, respectively. 
By allowing nonlinear dependences of the 'fluxes" on the "forces", one would leave the solid grounds of well-established statistical physics, since not very much is 
known on the validity range of such theories. Nevertheless, 
the relations (\ref{relax}, \ref{visc}) are nonlinear in the sense that the material tensors 
depend on the variables of the systems, in particular on temperature $T$, 
pressure and the strain tensor. The dependence on the scalar quantities 
is rather trivial and will not be shown explicitly, while the strain 
tensor dependence is more complicated. 
Starting from an equilibrium state $U_{ij}=0$ we can expand the material tensors into powers of $U_{ij}$. Up to quadratic order in 
Eqs.(\ref{udyn},\ref{vdyn}), which is what we need for the comparison in Chapt. IV, we find 
for the general form of the rank 4 material tensors
\begin{eqnarray} \label{alpha}
\alpha_{ijkl} &=&  \frac{ \alpha_1}{2} (\delta_{ik} \delta_{jl} 
+ \delta_{il} \delta_{jk}) + \frac{\alpha_2}{2} 
( U_{ik} \delta_{jl} + U_{jk} \delta_{il} 
+U_{il} \delta_{jk} +U_{jl} \delta_{ik} ) +
\alpha_3 \delta_{ij} \delta_{kl}  +  \nonumber \\ &&
\alpha_4 (\delta_{ij} U_{kl} + \delta_{kl} U_{ij}) 
+ \alpha_5\delta_{ij} \delta_{kl} U_{pp} 
+ \alpha_6 (\delta_{ik}\delta_{jl} + \delta_{il} \delta_{jk}) U_{pp}.
\end{eqnarray}
Since the "fluxes" are derived from a potential 
(the entropy production) by partial derivation, the symmetries 
of the tensors $A_{ij}$ and $\Psi_{ij}$ 
are transferred to $\sigma_{ij}^{(ph)}$ and 
$X_{ij}^{(ph)}$, respectively and are reflected in the material 
tensors being symmetric under the interchange of $i-j$, $k-l$, and $ij-kl$. 
For the viscosity tensor $\nu_{ijkl}$ we can simplify the 
form (\ref{alpha}), since $A_{kk}=0$. Thus, all parts 
$\sim \delta_{ij}$ or $\delta_{kl}$ are projected out in the 
entropy production and only the coefficients 
$\nu_{1,2,6}$ appear in Eq.(\ref{visc}).
Note that neither $U_{ij}$, nor $\Psi_{ij}$, 
nor the total stress tensor $\sigma_{ij}$ have to be traceless, 
despite $A_{ij}$ 
being so. If desired, 
the expansion (\ref{alpha}) can easily be continued to 
arbitrary order.  A more general form of Eqs.(\ref{relax},\ref{visc}) is 
 discussed in Appendix A. In the main part of this paper we will 
stick to the simplest form given above that constitutes the minimal generalization of
hydrodynamics, but still captures viscoelasticity.

The static phenomenological properties are expressed by a relation 
between the strain and the elastic stress
\begin{equation} \label{Hooke}
\Psi_{ij} = K_{ijkl} U_{kl}
\end{equation}
where the elastic tensor $K_{ijkl}$ generally depends on all variables, 
in particular on the strain tensor itself. 
In second order its form is slightly different from that of
the $\alpha$ tensor, since it is derived from an elastic energy, 
which is cubic in the strain tensor, 
resulting in the constraint $2K_6 = K_4$
\begin{eqnarray} \label{K}
K_{ijkl} &=&  \frac{ K_1}{2} (\delta_{ik} \delta_{jl} 
+ \delta_{il} \delta_{jk}) + \frac{K_2}{2} 
( U_{ik} \delta_{jl} + U_{jk} \delta_{il} 
+U_{il} \delta_{jk} +U_{jl} \delta_{ik} ) +
K_3 \delta_{ij} \delta_{kl}  +  \nonumber \\ &&
K_4 (\delta_{ij} U_{kl} + \delta_{kl} U_{ij} + \frac{1}{2} [\delta_{ik}\delta_{jl} + 
\delta_{il} \delta_{jk}]U_{pp} ) 
+ K_5\delta_{ij} \delta_{kl} U_{pp}.
\end{eqnarray}
Thus we are left with two 
linear elastic moduli $K_{1,3}$ and three quadratic ones $K_{2,4,5}$. 
To 
ensure positivity of the elastic energy, from which (\ref{Hooke}) is 
derived, third order terms with large and positive elastic 
moduli are necessary (see Appendix A).

\section{Stress tensor description \label{stressdescr}}

In the preceding section viscoelastic hydrodynamics was expressed 
by the strain tensor and its derivatives. We will now rewrite these 
equations with the stress tensor as variable by replacing the 
strain tensor (and its derivatives) by the stress tensor 
(and its derivatives). This can only be done in an approximate 
way, since the equations are nonlinear. We will set up a power 
series expansion up to second order in the (old and new) variables. 
Of course, the resulting equations are less general than the 
starting ones and only applicable, if quadratic nonlinearities 
are sufficient for the problem at hand. It is then also sufficient 
to restrict the phenomenological expansions 
(\ref{relax},\ref{visc},\ref{Hooke}) to quadratic order. 
In addition, we will neglect in this section also those 
phenomenological terms that are connected with the traces of 
the variables ($U_{kk}$, $\Psi_{kk}$, $X_{kk}^{(ph)}$), 
i.e. there are only six static and dynamic material parameters 
left ($\alpha_{1,2}$, $\nu_{1,2}$, and $K_{1,2}$). A more general 
treatment including the trace-related second order phenomenological 
constants will be given in Appendix B. 
With this proviso the dynamic strain equation reads
\begin{equation} \label{udyn2}
 \frac{d}{dt}U_{ij} - A_{ij} + U_{ki} \nabla_j v_k + U_{kj} \nabla_i v_k 
=  -\frac{1}{\tau_1} U_{ij} - \frac{1}{\tau_2} U_{ik} U_{jk}
\end{equation}
with the abbreviations $\tau_1^{-1} = \alpha_1 K_1$ and 
$\tau_2^{-1} = 2 \alpha_1 K_2 + 2 \alpha_2 K_1$. The stress tensor 
that enters the Navier-Stokes equation (\ref{vdyn})
is written as
\begin{equation} \label{stress2}
\sigma_{ij} = - K_1 U_{ij} + K_2' U_{ik} U_{jk} 
-\nu_1 A_{ij} - \nu_2 (U_{ik} A_{jk} + U_{jk} A_{ik})
\end{equation}
with $K_2' = 2 K_1 - 2 K_2$. 
Linearizing in the phenomenological expansions 
(\ref{relax},\ref{visc},\ref{Hooke}), i.e. $K_2 = \nu_2 = 
\tau_2^{-1} =0$, does not linearize the stress tensor, 
because of the convective nonlinearities in the strain relaxation 
equation (\ref{udyn2}). Only the condition $K_2' =0$ (and $\nu_2=0$) 
linearizes the stress tensor, which is however a rather artificial 
condition, since it requires the linear and quadratic elastic 
moduli to be equal exactly.

Taking the derivative $d/dt$ of $\sigma_{ij}$ in (\ref{stress2}) 
and replacing $dU_{ij}/dt$ according to Eq.(\ref{udyn2}) we get 
\begin{equation} \label{sigmadyn}
 \frac{d}{dt}\sigma_{ij} =
   - f[ U_{ij}, A_{ij}, \frac{d}{dt}A_{ij}, \Omega_{ij}]
\end{equation}
in terms of the strain tensor, the deformational flow and its derivative, 
and the vorticity $\Omega_{ij} 
= \tfrac{1}{2}(\nabla_j v_i - \nabla_i v_j)$ describing 
rotational flow. The latter enters 
due to the convective nonlinearities in (\ref{udyn2}). 
To convert this into the desired dynamic equation for the stress 
tensor we have to invert $\sigma_{ij}= \sigma_{ij} 
[U_{ij}, A_{ij}]$, Eq.(\ref{stress2}), 
into $U_{ij}=U_{ij}[\sigma_{ij}, A_{ij}]$. This is done 
approximately by the power expansion 
$U_{ij} = U_{ij}^{(lin)} + U_{ij}^{(quad)} 
+ \ldots$, where $U_{ij}^{(lin)}$ and $U_{ij}^{(quad)}$ 
contain expressions linear and quadratic in the variables, 
respectively. In particular we find
\begin{eqnarray} \label{ulin}
K_1 U_{ij}^{(lin)} &=& - \sigma_{ij} - \nu_1 A_{ij} \\ 
K_1^3 U_{ij}^{(quad)} &=& K_2' \sigma_{ik} \sigma_{jk} 
+(K_2' \nu_1 + K_1 \nu_2 )(\sigma_{ik} A_{jk} + \sigma_{jk} A_{ik}) 
+  \nu_1 (K_2' \nu_1 + 2 K_1 \nu_2 )A_{ik} A_{jk} \quad\quad
\label{uquad}
\end{eqnarray}
Using these expressions the dynamic equation for the stress 
tensor takes the final form 
\begin{eqnarray} 
{\tau_1} \frac{D_a}{Dt}\sigma_{ij}  + \sigma_{ij} &=&  
-\nu_{\infty} A_{ij} - {\nu_1}{\tau_1}   \frac{D_b}{Dt} A_{ij} 
+ \frac{r}{K_1}  \sigma_{ik} \sigma_{jk} \nonumber \\&& 
+ \frac{\tau_1 \nu_2}{ K_1} \left( [\sigma_{jk} 
+ \nu_1 A_{jk}] \frac{\partial}{\partial t} A_{ik} 
+ [\sigma_{ik} + \nu_1 A_{ik}] \frac{\partial}{\partial t} 
A_{jk}  \right) + O(3)
\label{sigmadyn2}
\end{eqnarray}
where $\nu_{\infty} = \nu_1 + \tau_1 K_1$ and $r = 
 \tau_1/\tau_2 - K_2'/K_1 $ and
\begin{equation} \label{defD}
\frac{D_a}{Dt} T_{ij} \equiv \frac{d}{dt} T_{ij} 
-a (T_{ik} A_{jk} + T_{jk} A_{ik}) 
- (T_{ik} \Omega_{jk} + T_{jk} \Omega_{ik}) 
\end{equation} 
for any tensor $T_{ij}$ and number $a$. For $a=-1$ ($a=+1$) 
$D_a/Dt$ is the lower (upper) convected derivative, for 
$a=0$ the Jaumann or corotational derivative, while for a 
general $a$ a linear combination of those is invoked. In 
our case the numbers $a$ and $b$ are
\begin{eqnarray}
\label{adef} a &=&  -1 + \frac {\nu_1}{K_1 \tau_2} 
 - \frac{K_2'}{K_1^2}
    \,  \frac{\nu_{\infty}}{ \tau_1}
\\ \label{bdef} b &=& -1  + \frac{\nu_1}{2K_1 \tau_2} 
- \frac{K_2'}{2K_1^2} \,\frac{ \nu_{\infty}}{\tau_1} - \frac{K_2'}{2K_1} - \frac{\nu_2}{\nu_1}
\end{eqnarray}
In the very special 'linear' case discussed above ($K_2'=\tau_2^{-1}=\nu_2=0$) 
we get 
$a=b=-1$ and both time derivatives 
in Eq.(\ref{sigmadyn2}) are of the lower convected type (as was the 
case for $U_{ij}$ without any approximation). However, 
this case is rather artificial and, generally, 
the phenomenological terms push these time derivatives away from 
the lower convected type and make them material dependent.

Splitting Eq.(\ref{sigmadyn2}) into the trace and the traceless part, 
it is easy to see that $\sigma_{kk}$ and $(d/dt)\sigma_{kk}$ are 
of O(2), since $A_{kk}=0$, and do not influence the dynamics of the 
traceless part $\sigma_{ij}^0$ up to second order.
This statement remains true, even if the trace-related second order 
phenomenological constants are taken into account (Appendix B).

\section{Comparison with constitutive models \label{compar}}

Eq.(\ref{sigmadyn2}) is in a form that many empirical constitutive 
models have and a direct comparison is possible. 
First there are the models of the form
\begin{eqnarray} 
{\tau_1} \frac{D_a}{Dt}\sigma_{ij}  
+ \sigma_{ij} &=&  -\nu_{\infty} A_{ij}  \label{sigmadynM}
\end{eqnarray}
that lack the time derivative of $A_{ij}$ 
and the terms $\sim r$ and $\sim \nu_2$. 
For $a=-1,0,1$ these are the Maxwell models \cite{bird}, 
while for $-1 < a < 1$ it is the Johnson-Segalman model \cite{JS}.
Apparently in these models the Newtonian viscosity 
$\nu_1$ has been neglected. For the actual viscosity 
$\nu_{\infty}$ that means it is given by $\tau_1 K_1$ alone.
This seems to be a good approximation for polymer solutions, 
where $\nu_1$ is interpreted as the viscosity of the solvent 
and  $\tau_1 K_1$ as due to the polymers. Of course, if  
$\nu_1=0$, it only makes sense to also put $\nu_2=0$. In addition, 
$r=0$ generally implies $\tau_2^{-1} = 0 = K_2'$ (the relation 
$\tau_1/\tau_2  = K_2'/K_1$ that also leads to $r=0$ is rather 
singular) meaning that these models are all within the linear case 
described above. However, this in turn has the consequence that 
$a=-1$, which makes the Johnson-Segalman 
model inconsistent. The Maxwell model with $a=1$ is applicable in a 
Lagrange description, while $a=0$ is not 
appropriate for a stress tensor like quantity (but rather for 
an orientational order parameter \cite{p2}). 

The next group of models take into account the time 
derivative of $A_{ij}$ \cite{oldroyd}
\begin{eqnarray} 
{\tau_1} \frac{D_a}{Dt}\sigma_{ij}  + \sigma_{ij} &=&  
-\nu_{\infty} A_{ij}  
- {\nu_{\infty}}{\lambda_r}   \frac{D_a}{Dt} A_{ij} \label{sigmadynO}
\end{eqnarray}
but neglect the terms $\sim r$ and $\sim \nu_2$. Thus, they again 
are in the linear class, a fact which is compatible with $a=b$ that 
is inherent to these models. Again $a=-1$ (Oldroyd A) is required 
for the Eulerian case (and $a=1$, or Oldroyd B, for the Lagrangian 
description) while $a=0$ (Jeffries) is inappropriate. 

The first model that goes beyond the linear case is 
the Giesekus model \cite{giesekus}
\begin{eqnarray} 
{\tau_1} \frac{D_1}{Dt}\sigma_{ij}  + \sigma_{ij} &=&  
-\nu_{\infty} A_{ij}    
-  \frac{2 c \tau_1}{\nu_{\infty}} \sigma_{ik} \sigma_{jk} \label{sigmadynG}
\end{eqnarray}
with the phenomenological number $c$ with $0<c<1$. This model 
neglects $\nu_1$ and consistently also $\nu_2$. Since $c$ is 
positive, comparison with (\ref{sigmadyn2}) shows that 
$r<0$ or $K_2'/K_1 > \tau_1/\tau_2$ is assumed. However, 
with $K_2' \neq 0$ one immediately concludes from (\ref{adef}) 
that $a$ has to deviate from the pure value $-1$ (or $1$). Thus, 
a model with nonlinearities in the stress tensor like the Giesekus 
model must not be of the pure lower (or upper) convected type, 
but has to show a material dependent deviation from that. 
For $\nu_1 = 0 = \nu_2$ this deviation from $a=-1$ is $-2c /K_1$. 

All the models above are restricted to nonlinearities quadratic 
in the variables and do not show any term we have hidden in 
$O(3)$ in Eq.(\ref{sigmadyn2}). This is different in the PTT and 
PT models \cite{PTT,PT} that have a stress tensor nonlinearity of 
the form $c \sigma_{ij} \sigma_{kk} +O(4)$, which for $A_{kk}=0$ 
is of third order. These models have consistently a stress convection 
with $a$ different from $-1$. However, there are several other 3rd 
order terms (cf. Eq.(\ref{sigmadynB})) and even 2nd order ones, 
not considered in these models.     

Another popular rheological model is the 
'second order fluid' \cite{bird} that contains a contribution
\begin{equation} \label{second}
\sigma_{ij}^{(2)} = -c_1 \frac{D_{\pm 1}}{Dt}  A_{ij} - c_2 A_{ik} A_{jk}
\end{equation}
in the stress tensor ($\pm$ for the Lagrangian and Eulerian picture, 
respectively).
Comparison with Eq.(\ref{stress2}) reveals that it is contained there 
with $c_1 = \nu_1 \tau_1$ and $c_2 = \nu_1 \tau_1 (b+1)$ 
(for the Eulerian case), where $b$ is defined in Eq.(\ref{bdef}). 
In the linear case defined above $b=-1$ and therefore $c_2=0$. Thus, 
the quadratic flow contributions to the stress tensor are due to the strain dependence  
of elasticity, viscosity and strain relaxation, when the strain field is taken into account as a relaxing variable and then eliminated.

Even when no convective derivatives are considered, very often 
nonlinear phenomenological stress/strain relations 
$\underline{{\sigma}}^{(ph)} = F(\underline{{A}})$ are used. As an 
example, in a power law fluid the Newtonian shear viscosity $\nu_1$ 
is replaced by $\nu = \sum_{n=1}^N \nu_n (A_{ik} A_{ik})^{(n-1)/2}$. 
Much more complicated forms are used \cite{boehme}.  The problem 
with all these models is the compliance with thermodynamics.  
The expansion (\ref{alpha}) for the viscosity tensor
in terms of $U_{ij}$ avoids this problem and can be carried on to 
any order desired. However, similar contributions to the stress tensor originate from 
the expansions of $\alpha_{ijkl}$ and $K_{ijkl}$. In addition, these expansions also change quite considerably the structure of the dynamic equation for the stress tensor (\ref{sigmadyn2}) rendering inconsistent any model that uses a power law description of the shear viscosity, only.

\section{Summary}

In this manuscript we have shown that the hydrodynamically derived model for non-Newtonian fluids in terms of the Eulerian strain tensor contains most of the standard rheo\-logical models as special cases and discards a few of them. Of course, the former is more general, as it contains powers of the relevant fields of arbitrary order when written in terms of the stress tensor. The hydrodynamic method allows to discriminate those parts of the dynamics that are due to general principles from the unavoidable phenomenological part. The latter is given here in the form of truncated power series in the strain tensor that can systematically be generalized when necessary.  For the phenomenological part we stick to the well-established 'linear irreversible thermodynamics', which, being linear in the generalized forces, nevertheless leads to equations highly nonlinear in the variables like the strain tensor.

\section*{}
ACKNOWLEDGEMENT - This research was supported in part by the 
National Science Foundation under Grant No. PHY99-07949.

\setcounter{equation}{0}
\renewcommand{\theequation}{A.\arabic{equation}}

\section*{APPENDIX A: Generalization of the 
phenomenological equations \label{appA}}

In Eqs.(\ref{relax},\ref{visc}) we have omitted possible 
reversible crosscouplings between flow and strain dynamics
\begin{eqnarray} \label{relaxg}
X_{ij}^{(ph)} &=&-\alpha_{ijkl} \Psi_{kl} - \beta_{ijkl} A_{kl}
\\ \label{viscg}
\sigma_{ij}^{(ph)} &=& - \nu_{ijkl} A_{kl} + \beta_{klij} \Psi_{kl}
\end{eqnarray}
characterized by the tensor $\beta_{ijkl}$, which is 
of a slightly more complicated  
form than $\alpha_{ijkl}$ in Eq.(\ref{alpha}). 
Being reversible it is not derived from a potential and 
lacks the $ij-kl$ symmetry. This results in two different 
components $\beta_{4a} \delta_{ij} U_{kl} + \beta_{4b} \delta_{kl} U_{ij}$, 
where, however, the latter (as well as $\beta_{3,5}$) vanish due 
to incompressibility. Thus we are left with 4 parameters $\beta_{1,2,4a,6}$.
Such terms, coming with 
probably small reactive transport parameters, have to compete with 
the parameter free, symmetry required terms already being part of 
Eqs.(\ref{udyn},\ref{vdyn}). These reversible crosscouplings are 
possible in the viscoelastic case only, and are absent for permanent 
elasticity.
In the latter case only diffusion $X_{ij}^{(ph)} 
= D \nabla_k (\nabla_i \Psi_{jk} + \nabla_j \Psi_{ik})$ is present.

We also list here the 5 third order static elastic contributions
that follow from an elastic energy quartic in the strain tensor
\begin{eqnarray} \label{elast3}
K_{ijkl}^{(3)} &=&
K_{6} \delta_{ij} \delta_{kl} U_{pp} U_{qq} 
+ K_{7} ([\delta_{ik} \delta_{jl} 
+ \delta_{il} \delta_{jk}] U_{pp} U_{qq} 
+  \delta_{ij} \delta_{kl}  U_{pq} U_{pq}) 
\\ 
&&+   K_8 (\delta_{ij} U_{kp} U_{lp} + \delta_{kl} U_{ip} U_{jp} 
+ \tfrac{1}{4} [\delta_{ik} U_{jl} + \delta_{jk} U_{il} 
+ \delta_{il} U_{jk} + \delta_{jl} U_{ik} ] U_{pp}) \nonumber \\ &&
+ K_{9} (\delta_{ik} \delta_{jl} 
+ \delta_{il} \delta_{jk}) U_{pq} U_{pq}
+ K_{10} ( \delta_{ik} U_{jp} U_{lp} 
+ \delta_{jk} U_{ip} U_{lp} + \delta_{il} U_{jp} U_{kp} 
+ \delta_{jl} U_{ip} U_{kp} ) \nonumber 
\end{eqnarray}
and that are a generalization of the first and second 
order terms kept in Sec.\ref{straindescr}.
A necessary stability condition involving the second order modulus $K_2$ requires 
$27 K_1 (K_9 + K_{10}) >2  K_2^2$.

\setcounter{equation}{0}
\renewcommand{\theequation}{B.\arabic{equation}}

\section*{APPENDIX B: Influence of the tensor traces
 \label{appB}}

Here we discuss the influence of those phenomenological parameters 
in Eqs.(\ref{alpha},\ref{Hooke}) that are connected with the traces 
of the tensors involved, 
$\alpha_{3,4,5,6}$, $\nu_6$, and $K_{3,4,5}$, as well as 
$\beta_{1,2,4a,6}$ (\ref{relaxg},\ref{viscg}) neglected in the main text.  
The phenomenological parts of the dynamic and static equations then read
\begin{eqnarray} 
X_{ij}^{(ph)} &=&-\alpha_1 \Psi_{ij} - \alpha_2 (U_{ik} \Psi_{jk} 
+ U_{jk} \Psi_{ik}) - \alpha_3 \delta_{ij} \Psi_{kk} 
- \alpha_4 (\delta_{ij} U_{kl} \Psi_{kl} + U_{ij} \Psi_{kk} )
-\alpha_5 \delta_{ij} U_{kk} \Psi_{ll}  \nonumber  \\
 && -2\alpha_6 U_{kk} \Psi_{ij}
 - \beta_1 A_{ij} - \beta_2 (U_{ik} A_{jk} + U_{jk} A_{ik} ) 
-\beta_{4a} \delta_{ij} U_{kl} A_{kl} - 2\beta_6 U_{kk} A_{ij}
\label{relaxB} \\ \label{viscB}
\sigma_{ij}^{(ph)} &=& - \nu_1 A_{ij} - \nu_2 (U_{ik} A_{jk} 
+ U_{jk} A_{ik})  - 2 \nu_6 A_{ij} U_{kk} + \beta_1 \Psi_{ij} 
+ \beta_2 (\Psi_{ik} U_{jk} + \Psi_{jk} U_{ik}) \nonumber  \\
 && + \beta_{4a} \Psi_{kk} U_{ij} + 2 \beta_6 \Psi_{ij} U_{ll}
\\ \label{HookeB}
\Psi_{ij} &=& K_1 U_{ij} + 2 K_2 U_{ik} U_{jk} + K_3 \delta_{ij} U_{kk}
+ K_4 (\delta_{ij} U_{kl} U_{kl} +2 U_{ij} U_{kk}) 
+ K_5 \delta_{ij} U_{kk} U_{ll} 
\end{eqnarray}
The strain relaxation and the stress take the form
\begin{eqnarray}
\frac{D_{-1}}{Dt} U_{ij} - A_{ij} &=& - \tau_1^{-1} U_{ij} 
- \tau_2^{-1} U_{ik} U_{jk} - \tau_3^{-1} \delta_{ij} U_{kk} 
- \tau_4^{-1} \delta_{ij} U_{kl} U_{kl} 
- \tau_5^{-1} \delta_{ij} U_{kk} U_{ll}   \\ && - \tau_6^{-1} U_{ij} U_{kk}  
 -\beta_1 A_{ij} 
- \beta_2 (U_{ik} A_{jk} + U_{jk} A_{ik}) \nonumber
-\beta_{4a} \delta_{ij} U_{kl} A_{kl} - 2 \beta_6 U_{kk} A_{ij}
\label{udynB} \\
\sigma_{ij} &=& - \tilde K_1 U_{ij} + \tilde K_2' U_{ik} U_{jk} 
- \tilde K_3 \delta_{ij} U_{kk} + \tilde K_3' U_{ij} U_{kk} 
- \tilde K_4 \delta_{ij} U_{kl} U_{kl}  \\&& \nonumber
- \tilde K_5' \delta_{ij} U_{kk} U_{ll} 
- \nu_1 A_{ij} - \nu_2 (U_{ik} A_{jk} + U_{jk} A_{ik} )
- 2 \nu_6 A_{ij} U_{kk}  \label{stressB}
\end{eqnarray} 
with the new parameters $\tau_3^{-1} = \alpha_1 K_3 + \alpha_3 K_1 
+ 3 \alpha_3 K_3$, $\tau_4^{-1} = \alpha_1 K_4 + 2 \alpha_3 K_2 
+ 3 \alpha_3 K_4 + \alpha_4 K_1$, $\tau_5^{-1} 
= 2\alpha_3 K_4 + \alpha_4 K_3 + \alpha_5 K_1 + 
\alpha_1 K_5 + 3 \alpha_3 K_5 + 3 \alpha_5 K_3 + 2 \alpha_6 K_3$, $\tau_6^{-1} 
= 2 \alpha_2 K_3 + \alpha_4 K_1 + 2\alpha_1 K_4 + 
3 \alpha_4 K_3+ 2 \alpha_6 K_1$, $\tilde K_1 
= K_1 (1-\beta_1)$, $\tilde K_3 = K_3 (1-\beta_1)$, $\tilde K_4 
= K_4 (1-\beta_1)$, $\tilde K_2' = 
K_2' +2 \beta_2 K_1 + 2 \beta_1 K_2$, 
 $\tilde K_3' = 2 \tilde K_3 (1+\beta_2 
+ \tfrac{3}{2} \beta_{4a}) -2 \tilde K_4 
+ K_1 (\beta_{4a} + 2 \beta_6)$, 
and $\tilde K_5' = K_5 (1-\beta_1) - 2 K_3 \beta_6$.

The procedure to switch from the strain relaxation equation to 
an effective dynamic equation for the stress tensor is the same 
as described in Sec. \ref{stressdescr}. Due to the many new terms, 
it is more involved, and it is complicated additionally by the 
fact that the traces and the traceless parts of 
$\sigma_{ij}$ and $U_{ij}$ are related by different 
parameters (even in the linear case) as can be seen from the 
modified Eq.(\ref{ulin})
\begin{equation}
\tilde K_1 U_{ij}^{(lin)} = - \sigma_{ij} - \nu_1 A_{ij} + 
\frac{\tilde K_3}{3 \tilde K_3 
+ \tilde K_1} \delta_{ij} \sigma_{kk} \label{ulinB}
\end{equation}
We refrain from writing down more details, but discuss the 
structure of the final result
\begin{eqnarray} \label{sigmadynB}
{\tau_1} \frac{D_{\tilde a}}{Dt}\sigma_{ij}  &
+& \sigma_{ij} + B_1 \delta_{ij} \sigma_{kk} 
= - \tilde \nu_{\infty} A_{ij} - {\nu_1}{\tau_1}   
\frac{D_{\tilde b}}{Dt} A_{ij} 
+ \frac{\tilde r}{\tilde K_1}  \sigma_{ik} \sigma_{jk}  +
B_2 \sigma_{kk} \frac{\partial}{\partial t} A_{ij} 
\\&+&  \frac{\tau_1 \nu_2}{ \tilde K_1} 
\left( [\sigma_{jk} + B_3 \delta_{jk} \sigma_{pp} 
+  \nu_1 A_{jk}] \frac{\partial}{\partial t} A_{ik} 
+ [\sigma_{ik} + B_3 \delta_{ik} \sigma_{qq}
+ \nu_1 A_{ik}] \frac{\partial}{\partial t} A_{jk}  \right) 
\nonumber \\ &+&  \sigma_{kk} (B_4 \sigma_{ij} + B_5  A_{ij}) 
+ \delta_{ij} (B_6 \sigma_{kl} \sigma_{kl} + B_7 \sigma_{kl} A_{kl} 
+ B_8 A_{kl} A_{kl} + B_9 \sigma_{kk} \sigma_{ll}) + O(3) \nonumber
\end{eqnarray}
with $\tilde \nu_{\infty} = \nu_1 + {K_1}{\tau_1}(1-\beta_1)^2 $ and 
the tilted numbers $\tilde a$, $\tilde b$, and $\tilde r$ being much 
more complicated than the untilted ones (\ref{adef},\ref{bdef}). 
There are structurally new terms related to $\sigma_{kk}$, a linear one 
(with $B_1 = \tilde K_3 / (3\tilde K_3 + \tilde K_1) + \tau_1/\tau_3)$, and eight quadratic ones
characterized by coefficients $B_{2,...,9}$. 
Nevertheless, the trace 
of the stress tensor and its time derivative are of second order and  
do not influence the dynamics of the traceless part in $O(2)$. 
Thus, for the deviator, $\sigma_{ij}^0 = \sigma_{ij} 
- (1/3) \delta_{ij} \sigma_{kk}$, the dynamic equation has exactly 
the form (\ref{sigmadyn2}), but with the tilted numbers and parameters 
instead of the untilted ones. For the trace we find to second order
\begin{eqnarray}
{\tau_1} \frac{d}{dt}\sigma_{kk}  + (1+ 3B_1)\sigma_{kk}  
&=&    {2 \tilde b \nu_1}{\tau_1} A_{kl} A_{kl} 
+ \frac{\tilde r}{\tilde K_1}  \sigma_{kl}^0 \sigma_{kl}^0 
+  \frac{2\tau_1 \nu_2}{ \tilde K_1} \left( \sigma_{kl}^0 
+  \nu_1 A_{kl} \right) \frac{\partial}{\partial t} A_{kl} 
\nonumber \\ &+&  3 ( B_6 \sigma_{kl}^0 \sigma_{kl}^0 
+  [B_7 + \frac{2 }{3 }\tau_1 \tilde a ]\, \sigma_{kl}^0 A_{kl} 
+  B_8 A_{kl} A_{kl}) + O(3) \quad \label{tracedynB}
\end{eqnarray}
indicating that its relaxational dynamics is completely given by 
the flow $A_{ij}$ and the traceless part $\sigma_{ij}^0$.

\end{document}